
\font\twelvebf=cmbx12
\font\ninerm=cmr9
\nopagenumbers
\magnification =\magstep 1
\overfullrule=0pt
\baselineskip=18pt
\line{\hfil CCNY-HEP 4/94}
\line{\hfil May 1994}
\vskip .8in
\centerline{\twelvebf Hard Thermal Loops in a Moving Plasma and }
\centerline{\twelvebf a Magnetic Mass Term}
\vskip .5in
\centerline{\ninerm V.P. NAIR}
\vskip .1in
\centerline{ Physics Department}
\centerline{City College of the City University of New York}
\centerline{New York, New York 10031.}
\centerline{E-mail: vpn@ajanta.sci.ccny.cuny.edu }
\vskip 1in
\baselineskip=16pt
\centerline{\bf Abstract}
\vskip .1in
We consider the hard thermal loops of Quantum Chromodynamics
for a moving quark-gluon plasma. Generalizing from this we
suggest a candidate for the magnetic mass term. This mass term may also
be useful in understanding the mass gap of three-dimensional
non-Abelian gauge theories.
\vfill\eject
\footline={\hss\tenrm\folio\hss}
\def \12 {{\textstyle {1\over 2}}}

\def \vk {{\vec k}}
\def \vQ {{\vec Q}}
\def \bz {{\bar z}}
\def \bu {{\bar u}}
\def \bv {{\bar v}}
\def \bw {{\bar w}}
\def \bn {{\bar n}}
\def \dA {{\dot A}}
\def \dB {{\dot B}}
\def \tA {{\tilde A}}
\def \tG {{\tilde{\Gamma}}[A]}

\magnification =\magstep 1
\overfullrule=0pt
\baselineskip=20pt
\pageno=2
Hard thermal loops in Quantum Chromodynamics (QCD) at high temperatures have
been
the subject of many recent investigations [1-6]. The generating functional
$\Gamma [A]$
of the hard thermal loops is a gauge-invariant nonlocal functional of the gauge
potential $A_\mu$ and is essential in describing Debye screening and Landau
damping
effects as well as in carrying out the Braaten-Pisarski
reorganization of perturbation theory [1,2,3].
$\Gamma [A]$ also has an elegant mathematical description,
being closely related to
the eikonal for a Chern-Simons (CS) theory [2].

As mentioned above, $\Gamma [A]$ is essentially an electric mass term for
gluons and
properly incorporating $\Gamma [A]$ in any calculation eliminates some of the
infrared singularities. There would still remain some singularities since the
static magnetic interactions are not screened. It is generally believed that
for QCD at high temperatures there is also a magnetic mass term which screens
the static magnetic interactions (or more generally magnetic fields with
spacelike
momenta). One way to understand how this might happen is as follows [7]. In the
standard imaginary-time formalism, the partition function of QCD can be written
as
a Euclidean path integral with the fields periodic in the time-interval $[0,
\beta =
(1/T) ~]$, $T$ being the temperature.
At high temperatures and for wavelengths long compared to $\beta$, the
modes with nonzero Matsubara frequencies are unimportant and the theory reduces
to three-dimensional QCD with a coupling constant ${\sqrt{g^2 T}}$, $g$ being
the
coupling constant of the four-dimensional theory. For QCD in three dimensions
we expect a mass gap ($\sim g^2T$) and this is effectively the magnetic mass
of the high temperature four-dimensional QCD. While this argument is
suggestive,
it is by no means adequate and understanding the structure of the magnetic mass
term and the calculation of its value have been difficult, although many
approaches have been tried [7,8].

In this paper, we propose a candidate magnetic mass term. Of course, such a
mass term
must be gauge-invariant and parity-even. In the rest frame of the plasma, which
is usually
used for temperature-dependent calculations, it will not be Lorentz-invariant.
However, if we include the overall motion of the plasma, we must
have Lorentz invariance as well. We should have Lorentz invariance in this
qualified sense; a mass term which is Lorentz-invariant, independently of the
motion of the plasma is not acceptable. The effective fields which exhibit
local
behaviour at long wavelengths are presumably related to the gluon fields by
nonlocal transformations and so locality is not a priori necessary for the mass
term. (Recall that the electric mass term $\Gamma [A]$ is anyway nonlocal.)
Finally the putative mass term must be essentially three-dimensional in
consistency
with the dimensional reduction argument sketched above; i.e., it should give a
mass
only to the spatial components of $A_\mu $ and in a rotationally invariant way.
We try to construct a magnetic mass term based on these requirements. Our
strategy
is to start from hard thermal loops or $\Gamma [A]$. We first generalize it to
a moving plasma. It is then easy to see, from the structure of this generalized
$\Gamma [A]$, that there is another gauge-invariant mass term, which we denote
by ${\tilde{\Gamma}}[A]$, which is also Lorentz-invariant in the qualified
sense described above. Evaluating $\tG$ in the rest frame of the plasma,
 we see that
it does indeed screen static magnetic interactions. $\tG$ is our candidate
magnetic mass term. (A mass term which is not parity-conserving has been
analyzed in ref. [9]; however it does not seem to be applicable to our
problem.)

While $\tG$ may give the right structure for the magnetic mass, we have no
computation
of its coefficient or the value of the magnetic mass. Nevertheless, even just
understanding the structure of the magnetic mass can be useful. At the very
least,
$\tG$ gives a mathematically precise and gauge-invariant way of introducing an
infrared
cutoff in calculations of processes in the quark-gluon plasma. It may also be
possible,
knowing $\tG$, to set up a gauge-invariant self-consistent evaluation of the
magnetic mass.

The generating functional for hard thermal loops can be written as
$$
\Gamma [A]~= m^2 \int d \Omega~ K[A_+, A_- ] \eqno(1)
$$
where $A_+= \12 A \cdot Q,~A_-= \12 A\cdot Q',~ Q^\mu= (1, \vQ),~Q'^\mu
=(1, -\vQ)$.
$A_\mu= A_\mu^a (-it^a)$ where $t^a$ are hermitian matrices giving a basis for
the
fundamental representation of the
Lie algebra of the gauge group, here taken to be $SU(N)$. Further $\vQ^2=1$,
so that $Q^\mu $ and $Q'^\mu $ are null vectors, i.e.,
$Q^\mu Q_\mu = Q'^\mu Q'_\mu =0$. The $d\Omega$-integration in (1) is over
the
orientations of the unit vector $\vQ$.
In the lowest order analysis, $m^2= (N+ \12 N_F){T^2\over 6}$ where $N$ is
the number of
colors and $N_F$ is the number of quark-flavors.
We also have
$$
K[A_z, A_\bz]= -{1\over \pi}\int d^2x^T \left[ ~\int d^2z~{\rm Tr}
(A_z A_{\bar z}) ~+
i\pi I(A_z) ~+ i\pi I(A_{\bar z}) \right] \eqno(2)
$$
where $x^T$ is transverse to $\vQ $, i.e. $x^T \cdot Q =0$. Also
$$
I(A_z) = i\sum_2^{\infty}  {(-1)^n\over n} \int {d^2z_1\over \pi}...
{d^2z_n\over \pi} ~{{{\rm Tr}(A_z(x_1) \cdots A_z(x_n)) \over {{\bar z}_{12}
\cdots {\bar z}_{n-1 n}{\bar z}_{n1} }}}
\eqno(3)
$$
Here ${\bar z}_{ij}= {\bar z}_i -{\bar z}_j$ and all $A$'s in (3) have the same
argument for the transverse coordinates $x^T$. We have written down the
Euclidean version; $K[A_+,A_-]$ and $I(A_+)$ are obtained by a
simple continuation. $z$ and $\bar z$
denote the Euclidean version of the
lightcone coordinates $(Q'\cdot x),~ (Q\cdot x)$ respectively;
correspondingly $A_+ $ and $A_-$ have the
Euclidean versions $A_z,~A_{\bar z}$. The Minkowski version is obtained by
noting that ${1\over \pi }{1\over {\bz -\bz'}}$ is the Green's function for
$\partial_z$ and continues to the Green's function
$$
G(x-y) =~2i\int {d^4k \over (2\pi)^4} e^{-ik(x-y)} ~{1\over k\cdot Q}. \eqno(4)
$$
$I(A_-)$, or its Euclidean counterpart $I(A_\bz )$, is given by (3) with
$Q \leftrightarrow Q'$.  $I(A_z)$ is related to the
eikonal for a CS theory. It is also possible to write $K[A_+, A_-]$ in
terms of the action for a Wess-Zumino-Novikov-Witten (WZNW) theory [10].

We must now generalize (1) to a moving plasma. One can derive $\Gamma [A]$
in a moving frame by recalculating it with the appropriately
generalized statistical distribution functions [11]. But it is simpler to
obtain it by just using Lorentz invariance. Once we make a
Lorentz transformation, the time-component of
$Q_\mu$ is no longer $1$, so we need to consider more general null vectors.
A null vector can generally be written as the product of two spinors,
viz. as $ u^A e^\mu_{A\dA} \bu^{\dA}$, where $e^\mu= (1,\sigma^i),~\sigma^i$
being the Pauli matrices and $u^A$ is a two-spinor, with $\bu^{\dA}=
{\overline {(u^A)}}$, the complex conjugate of $u^A$. The Lorentz-invariant (or
$SL(2,{\bf C})$-invariant) tensors are ${\epsilon}^{AB}$ and
$\epsilon^{\dA \dB}$, with, of course, no mixing of
the dotted and undotted indices. These can be used to form invariant scalar
products of spinors. Although not essential, this
spinor notation will be quite useful in what follows.

Rather than giving a step-by-step generalization, it is simpler to write
down the manifestly Lorentz-invariant version of (1) and show that it
reduces to (1) in the rest frame of the plasma. Let $p^\mu$ be the overall
velocity vector of the plasma. The appropriate generalization of (1) is
given by
$$
\Gamma [A]= 2m^2 i\int d\mu ~~\Delta (u,v)~ {{K(A_u, A_v) }\over
{(u\cdot v)(\bu \cdot \bv)}}  \eqno(5)
$$
where we have two sets of two-component spinors $(u^A, \bu^{\dA})$ and
$(v^A,\bv^{\dA})$ (see Eq. (9) for an explicit parametrization) and
$$
A_u = \12 u^A(A\cdot e)_{A\dA}\bu^{\dA},~~~~~~~~~~
A_v= \12 v^A(A\cdot e)_{A\dA}\bv^{\dA}, \eqno(6)
$$
$$
d\mu ={{u\cdot du ~~\bu \cdot d\bu ~~v\cdot dv ~~\bv\cdot d\bv}\over {(u\cdot
v)^2
(\bu \cdot \bv)^2}}\eqno(7)
$$
$$
\Delta (u,v)= (u\cdot v) (\bu \cdot \bv) ~\delta (v(p\cdot e)\bu )
{}~\delta (u(p\cdot e)\bv) \eqno(8)
$$
Here $u\cdot du = u^1 du^2 -u^2 du^1$, etc.
The measure $d\mu $ and the generalized $\delta$-function $\Delta (u,v)$ are
invariant under complex rescalings of $u,v$, i.e., under $u\rightarrow
\lambda_1 u,~v\rightarrow \lambda_2 v$. The space of spinors is thus
two copies of ${\bf C}P^1=S^2$ with an identification between the two
enforced by the $\delta$-functions. This identification depends on $p^\mu$
and hence is frame-dependent. Expression (5) is manifestly Lorentz-invariant
if one transforms $p^\mu$ as well.

We now evaluate (5) in the rest frame of the plasma. Introduce
the parametrization
$$
\eqalign{
u &= \rho \left( \matrix{1\cr
z\cr}\right)
{}~~~~~~~~~~\bu = {\bar \rho} \left( \matrix{1\cr
                                            \bz \cr}\right) \cr
v &=\sigma \left(\matrix{\bw \cr
-1\cr}\right) ~~~~~~~~\bv = {\bar \sigma}
\left(\matrix{w\cr
-1\cr}\right) \cr} \eqno(9)
$$
We then have
$$
d\mu = {{dz~d\bz~dw~d\bw }\over {(1+z\bw )^2 (1+\bz w)^2}}. \eqno(10)
$$
With $p^\mu =(1,0,0,0)$, the $\Delta$-function becomes
$ \Delta (u,v)= (1+\bz w)(1+z \bw ) \delta (w-z)\delta (\bw -\bz)$.
Further
$
A_u = \12 (A\cdot Q) ~\rho {\bar \rho} ~(1+z \bz ),~~A_v= \12
(A\cdot Q') ~\sigma {\bar \sigma} ~(1+w \bw )
$
where
$$
Q^\mu = (1, ~{{z+\bz }\over {(1+z\bz )}}, ~{{i(z-\bz )}\over{(1+z\bz )}},
{}~{{1-z\bz }\over {(1+z\bz )}}) \equiv (1, ~\vQ (z))\eqno(11)
$$
and $Q'^\mu =(1, -\vQ (w))$. The standard parametrization of $\vQ$ in
terms of the direction cosines is obtained if we write $z= e^{-i\varphi} \tan
(\theta /2)$. $\rho, \sigma$-dependence cancels out since $K$ is
homogeneous of degree two in these variables. Integrating out the
$\delta$-functions, we see
that (5) indeed becomes (1) in the rest frame. One can in a straightforward
manner evaluate the $n$-point functions in an arbitrary frame; the expressions,
even for the two-point function, are too long to be displayed here.

We now consider alternatives to (5). The possible changes
are in the $\delta$-functions or in the combinations $A_u, ~A_v$.
For the argument of the $\delta$-functions, we can use $v(p\cdot e)\bu $
as we have done or $v(p\cdot e)\bv $ (and their conjugates). The second choice
implies the vanishing of $v^A$ in the rest frame and is not acceptable. The
only modification is thus in the combinations for the gauge potential. We can
have (6) or
$$
{\tA}_u= \12 u^A (A\cdot e)_{A \dA}\bv^{\dA},~~~~~~~~~~\tA_v =
\12 v^A (A\cdot e)_{A\dA}\bu^{\dA} \eqno(12)
$$
We thus consider the term
$$
\tG = (-2M^2 i) \int d\mu~~ \Delta (u,v)~{{K[\tA_u, \tA_v]}\over {(u\cdot v)
(\bu \cdot \bv )}} \eqno(13)
$$
In the rest frame of the plasma this simplifies to
$$
\tG = - M^2 \int d\Omega~ K[A_n, A_{\bn}] \eqno(14)
$$
where $A_n= \12 A_in_i,~ A_{\bn}= \12 A_i\bn_i$ and
$
n_i= (-\cos \theta \cos \varphi -i \sin \varphi,~-\cos \theta
 \sin \varphi +i \cos \varphi,~ \sin \theta~)
$,
using $z=e^{-i\varphi} \tan (\theta /2)$. Notice that $n_i$ is a complex
three-dimensional null vector. $\tG$ involves, in the rest frame, only the
spatial components of $A_\mu$ as expected for a magnetic mass term.
$I(A_n),~I(A_{\bn})$ are defined using $n\cdot x$ and $\bn \cdot x$
in place of $z,\bz$ in (3). Thus $K$ in (14) is $K[\12 A\cdot Q, \12
A\cdot Q' ]$ of (2,3) with $Q^\mu\rightarrow (0,n_i),~Q'^\mu \rightarrow
(0,\bn_i )$.

Consider the simplification of the quadratic term in (14). Using
$\int d\Omega~ n_i \bn_j = {8\pi \over 3} \delta_{ij}$ and
$$
\int d\Omega ~{k\cdot \bn \over {k\cdot n}} n_i n_j= {8\pi \over 3}
\left[ {k_ik_j \over \vk^2}-{1\over 2} \left(\delta_{ij}-
{k_ik_j \over \vk^2}\right)\right] \eqno(15)
$$
we find
$$
\tG = -{M^2 \over 2}\int {d^4k \over (2\pi )^4}~ A^a_i (-k) \left( \delta_{ij}-
{k_ik_j\over \vk^2}\right) A^a_j(k) ~+ {\cal O}(A^3) \eqno(16)
$$
Thus $\tG$ does give screening of transverse magnetic interactions, with a
screening mass $M$.

The higher order
terms in $\tG$ can also be evaluated in a straightforward fashion, noting
that the basic change is replacing $Q^\mu$ by $(0,n_i) $ and $Q'^\mu $
by $(0,\bn_i )$. ( Expressions in $n_i, \bn_i$ which arise from the
projectively
invariant spinorial expressions lead to rotationally invariant results.
As a word of caution, an arbitrary expression involving the $n_i,\bn_i$
cannot be written in the spinor language without explicitly breaking rotational
symmetry. Such a complication, however, does not arise for $\tG$.) Notice also
that $\tG$
is explicitly real, eventhough $A_n,~A_{\bn}$ are complex even in Minkowski
space. In writing out $\tG$, we need the inverses of $n\cdot \partial$ and
$\bn \cdot \partial$. Essentially only the spatial coordinates are involved
and the boundary condition is the vanishing of the Green's function at
infinity. The derivatives $n\cdot \partial,~\bn\cdot \partial$ behave
like $\partial_z,~\partial_{\bz}$ with Green's functions which are the
analogues of $(\bz -\bz')^{-1},~(z-z')^{-1}$. In particular,
we do not need an $i\epsilon$-prescription to define the inverses.

The various terms in $\tG$ are clearly nonlocal. Just as in the case
of hard thermal loops [3,4], it is possible to introduce
auxiliary field variables and write the equations of motion in a local
way. The equations of motion can be written as
$$
\eqalignno{
(D_\mu F^{\mu\nu})^a ~-J^{\nu a}~&=0 &(17a)\cr
\partial_+ {\cal A}_- ~-\partial_- A_+ ~+[A_+, {\cal A}_-] &=0&(17b)\cr}
$$
$$
\eqalign{J^{\nu a}&= {M^2\over 2\pi}\int d\Omega~ {\rm Tr}\left\{ (-it^a)[
({\cal A}_- -A_-)Q^\nu + ({\cal A}_+ -A_+)Q'^\nu ]\right\}\cr
&={M^2\over 2\pi} \int d\Omega~{\rm Tr}\left\{ (-it^a)[ H^{-1}D_-H ~Q^\nu -
D_+H~H^{-1} Q'^\nu]\right\}\cr}\eqno(18)
$$
where $D_\mu$ denotes the covariant derivative and ${\cal A}_+= H A_+ H^{-1}
-\partial_+H H^{-1},~{\cal A}_-= H^{-1}A_- H+H^{-1}\partial_- H$. $H$ is
the auxiliary
matrix field, which is hermitian in the present case.
These equations have the same form
as the equations with the hard thermal loop contributions;
the difference is that $Q^\mu,~Q'^\mu$ are now given by $(0,n_i),~(0, \bn_i)$.

We can also write $n\cdot A = -n\cdot \partial U~U^{-1},~\bn\cdot A=
U^{\dagger -1}\bn\cdot \partial U^\dagger$ where $U$ is an
$SL(N,{\bf C})$-matrix. In this parametrization, $\tG$ becomes
$$
\tG= M^2 \int d\Omega ~dx^0 dx^T~S_{WZNW}(G) \eqno(19)
$$
where
$$
 S_{WZNW}(G)~=~-{1\over 2\pi}\int_{{\cal M}^2} d^2x~ {\rm Tr}(\partial_{+}G
G^{-1}\partial_{-}G G^{-1} ) ~+~ {1\over 12\pi} \int_{{\cal M}^3} {\rm Tr}(dG
G^{-1})^3 \eqno(20)
$$
In Eq.(19), $G$ is the hermitian matrix $U^\dagger U$ and
the integration is over the directions transverse to $n_i, \bn_i$, i.e., over
$x^0$ and the spatial direction $x^T$.
Notice that we can also write
$$
\eqalign{
{1\over 4}F^a_{ij}F^a_{ij}&= -{3\over 4\pi}\int d\Omega~
{\rm Tr}\bigl(F_{n\bn}F_{\bn n}\bigr)\cr
&= -{3\over 4\pi}\int d\Omega~{\rm Tr}\left[ \partial_{\bn} (\partial_n
G~G^{-1})
\partial_n ( \partial_{\bn}G~G^{-1})\right]\cr}\eqno(21)
$$

This result also shows that the action for (Euclidean)
three-dimensional
gauge theory, with the mass term added, can be written as
$$
\eqalign{
{\cal S}^{(3)}&= \int d\Omega ~dx^T~ {\cal S}^{(2)}\cr
{\cal S}^{(2)}&= -{3\over 4\pi}\int d^2z~
{\rm Tr}\bigl(F_{n\bn}F_{\bn n}\bigr)~- M^2 S_{WZNW}(G) \cr}
\eqno(22)
$$
${\cal S}^{(2)}$ is the action for two-dimensional QCD with an extra
WZNW-action; the fields do depend on all three coordinates but the
transverse coordinate $x^T$ only plays the role of a parameter
as far as ${\cal S}^{(2)}$ is concerned. This representation may
be useful in understanding the mass gap in three-dimensional gauge theories.
Of course, as we have mentioned before, we do not have a calculation of
$M^2$, either for QCD at finite temperature or intrinsically for the
three-dimensional gauge theory. In principle, one should not have to add
such a mass term, it should emerge from a calculation of the effective
action. A possible alternative, knowing the structure of this mass term,
is to seek a self-consistent evaluation of $M^2$. In the context of the
magnetic
mass for the QCD plasma, such an approach has been attempted in reference [8].
With our mass term, this calculation can perhaps be refined,
especially regarding questions of gauge invariance.

I thank R.Jackiw and D.Minic for useful discussions.
\vskip .2in
\noindent {\bf References}
\vskip .2in
\item{[1]} R. Pisarski, {\it Physica} {\bf A 158}, 246 (1989);
{\it Phys.Rev.Lett.}
{\bf 63}, 1129 (1989); E. Braaten and R. Pisarski, {\it Phys.Rev.}
{\bf D 42}, 2156 (1990); {\it Nucl.Phys.} {\bf B 337}, 569 (1990);
{\it ibid.} {\bf B 339},
310 (1990); {\it Phys.Rev.} {\bf D 45}, 1827 (1992);
J. Frenkel and J.C. Taylor, {\it Nucl.Phys.} {\bf B 334}, 199 (1990);
J.C. Taylor and S.M.H. Wong, {\it Nucl.Phys.} {\bf B 346}, 115 (1990).
\vskip .1in
\item{[2]} R. Efraty and V.P. Nair, {\it Phys.Rev.Lett.} {\bf 68}, 2891 (1992);
{\it Phys.Rev.} {\bf D 47}, 5601 (1993).
\vskip .1in
\item{[3]} R. Jackiw and V.P. Nair, {\it Phys.Rev.} {\bf D 48}, 4991 (1993)
\vskip .1in
\item{[4]} V.P. Nair, {\it Phys.Rev.} {\bf D 48}, 3432 (1993); CCNY Preprint
2/94, March 1994 (to appear in  {\it Phys. Rev.D}).
\vskip .1in
\item{[5]}J.P. Blaizot and E. Iancu, {\it Phys.Rev.Lett.} {\bf 70}, 3376
(1993); Saclay Preprint T93/064 (to appear in {\it Nucl.Phys.} B);
Saclay Preprints T94/02,03, 013 (1994).
\vskip .1in
\item{[6]} R. Jackiw, Q. Liu and C. Lucchesi, MIT preprint CTP\#2261
(1993); P.F. Kelly, Q. Liu, C. Lucchesi and C. Manuel, MIT Preprint
CTP\#2292, March 1994.
\vskip .1in
\item{[7]} A.D. Linde, {\it Phys.Lett.} {\bf B 96}, 289 (1980);
D. Gross, R. Pisarski and L. Yaffe, {\it Rev.Mod.Phys.} {\bf 53}, 43 (1981).
\vskip .1in
\item{[8]} O.K. Kalashnikov, {\it JETP Lett.}  {\bf 39}, 405 (1984).
\vskip .1in
\item{[9]} S.M. Carroll, G.B. Field and R. Jackiw, {\it Phys.Rev.} {\bf D 41},
 1231 (1990).
\vskip .1in
\item{[10]} S.P. Novikov, {\it Usp.Mat.Nauk} {\bf 37}, 3 (1982); E. Witten,
{\it Commun.Math.Phys.} {\bf 92}, 455 (1984).
\vskip .1in
\item{[11]}  S.R. de Groot, W.A. van Leeuwen and C.G. van Weert,
{\it Relativistic
Kinetic Theory: Principles and Applications}, North Holland, New York (1980).
\end